\documentclass{ws-procs9x6}
\usepackage{cite,feynarts,graphicx,amsmath,wrapfig}
\makeatletter
\def\citer{\@ifnextchar [{\@tempswatrue\@citexr}{\@tempswafalse\@citexr[]}}
 
\def\@citexr[#1]#2{\if@filesw\immediate\write\@auxout{\string\citation{#2}}\fi
  \def\@citea{}\@cite{\@for\@citeb:=#2\do
    {\@citea\def\@citea{--\penalty\@m}\@ifundefined
       {b@\@citeb}{{\bf ?}\@warning
       {Citation `\@citeb' on page \thepage \space undefined}}%
\hbox{\csname b@\@citeb\endcsname}}}{#1}}
\makeatother

\newcommand{\cal}{}

\def\reffi#1{\mbox{Fig.~\ref{#1}}}

\def\citere#1{\mbox{Ref.~\cite{#1}}}

\def\xtilde#1{%
  \setbox0\hbox{$\tilde#1$}%
  \rlap{\raise\ht0\hbox{\tiny$_{\,(\;\,)}$}}%
  \tilde#1%
}

\def\order#1{${\cal O}(#1)$}

\newcommand{\onel}{one-loop}

\newcommand{\MW}{M_W}
\newcommand{\MZ}{M_Z}
\newcommand{\MA}{M_A}

\newcommand{\MH}{M_H}

\newcommand{\MHp}{M_{H^\pm}}

\newcommand{\tsf}{\theta\kern-.20em_{\tilde{f}}}
\newcommand{\tsfp}{\theta\kern-.20em_{\tilde{f}\prime}}
\newcommand{\tsq}{\theta\kern-.15em_{\tilde{q}}}

\newcommand{\VL}{\left( \begin{array}{c}}
\newcommand{\VR}{\end{array} \right)}
\newcommand{\ML}{\left( \begin{array}{cc}}
\newcommand{\MLd}{\left( \begin{array}{ccc}}
\newcommand{\MLv}{\left( \begin{array}{cccc}}
\newcommand{\MR}{\end{array} \right)}

\newcommand{\tb}{\tan \beta}

\newcommand{\gev}{\,\, {\rm GeV}}

\newcommand{\BC}{\begin{center}}
\newcommand{\EC}{\end{center}}
\newcommand{\BE}{\begin{equation}}
\newcommand{\EE}{\end{equation}}
\newcommand{\BEA}{\begin{eqnarray}}
\newcommand{\BEAnn}{\begin{eqnarray*}}
\newcommand{\EEA}{\end{eqnarray}}
\newcommand{\EEAnn}{\end{eqnarray*}}

\newcommand{\id}{{\rm 1\kern-.12em
\rule{0.3pt}{1.5ex}\raisebox{0.0ex}{\rule{0.1em}{0.3pt}}}}

\def\al{\alpha}

\def\ga{\gamma}

\def\de{\delta}

\def\Si{\Sigma}

\newcommand{\eennH}{$e^+e^- \to \bar\nu \nu \, H$}
\newcommand{\eeenH}{$e^+e^- \to e \nu_e \, H^\pm$}

\newcommand{\eeenHm}{$e^+e^- \to e^+ \nu_e \, H^-$}

\newcommand{\fb}{\mbox{~fb}}

\allowdisplaybreaks

\def\limfunc#1{\mathop{\mathrm{#1}}}

\begin{document}

\thispagestyle{empty}
\setcounter{page}{0}
\def\thefootnote{\fnsymbol{footnote}}

\begin{flushright}
CERN--PH--TH/2004--021 \hfill
DCPT/04/06\\ 
IPPP/04/03 \hfill
MPI--PhT/2004--18\\
PITHA--04--02 \hfill
hep-ph/0402053\\
\date{\today}
\end{flushright}

\vspace{0.5cm}

\begin{center}

{\large\sc {\bf Single Charged MSSM Higgs-boson production }}

\vspace*{0.3cm} 

{\large\sc {\bf at a Linear Collider}}
\footnote{Talk presented by S.~Heinemeyer at {\it SUSY 2003:
Supersymmetry in the Desert}, held at the University of Arizona,
Tucson, AZ, June~5-10, 2003. To appear in the Proceedings.}

\vspace{0.5cm}

{\sc O.~Brein$^{1\,}$%
\footnote{
email: Oliver.Brein@physik.rwth-aachen.de
}%
, T.~Hahn$^{\,2}$%
\footnote{
email: hahn@feynarts.de
}%
, S.~Heinemeyer$^{\,3}$%
\footnote{
email: Sven.Heinemeyer@cern.ch
}%
~and G.~Weiglein$^{\,4}$%
\footnote{
email: Georg.Weiglein@durham.ac.uk
}%
}

\vspace*{0.5cm}

$^1$ Institut f\"ur Theoretische Physik E, RWTH Aachen, 
D--52056 Aachen, Germany

\vspace*{0.3cm}

$^2$ Max-Planck-Institut f\"ur Physik,
F\"ohringer Ring 6, D--80805 Munich, Germany

\vspace*{0.3cm}

$^3$ Dept.\ of Physics, CERN, 
TH Division,
1211 Geneva 23, Switzerland

\vspace*{0.4cm}

$^4$ IPPP, University of
  Durham, Durham DH1~3LE, UK

\end{center}

\vspace*{0.5cm}

\begin{center}
{\bf Abstract}
\end{center}
In the Minimal Supersymmetric Standard Model we present the calculation
of the single charged Higgs-boson production in the $\ga W$- or
$ZW$-fusion and the charged Higgs strahlung channel, \eeenH. The 
set of all \order{\al} corrections arising from loops of Standard
Model fermions and
scalar fermions are taken into account.  Contrary to the case of single
neutral heavy CP-even Higgs-boson production, for the charged Higgs
boson we find for all the parameter space of the typical benchmark
scenarios a cross section smaller than $\sim 0.01 \fb$ for 
$\sqrt{s}/2 \lesssim \MHp$. 

\def\thefootnote{\arabic{footnote}}
\setcounter{footnote}{0}

\newpage

\title{Single Charged MSSM Higgs-boson production at a 
Linear Collider%
\footnote{\uppercase{T}alk presented by \uppercase{S}. \uppercase{H}einemeyer
at {\it \uppercase{SUSY} 2003:
\uppercase{S}upersymmetry in the \uppercase{D}esert}\/, 
held at the \uppercase{U}niversity of \uppercase{A}rizona,
\uppercase{T}ucson, \uppercase{AZ}, \uppercase{J}une 5-10, 2003.
\uppercase{T}o appear in the \uppercase{P}roceedings.}}

\author{OLIVER BREIN}

\address{Institut f\"ur Theoretische Physik E, RWTH Aachen, 
D--52056 Aachen, Germany\\
E-mail: Oliver.Brein@physik.rwth-aachen.de}

\author{THOMAS HAHN}

\address{Max-Planck-Institut f\"ur Physik,
F\"ohringer Ring 6, D--80805 Munich, Germany\\
E-mail: hahn@feynarts.de}

\author{SVEN HEINEMEYER}

\address{Department of Physics, CERN, 
Theory Division,
1211 Geneva 23, Switzerland\\
E-mail: Sven.Heinemeyer@cern.ch}

\author{GEORG WEIGLEIN}

\address{IPPP, University of
  Durham, Durham DH1~3LE, UK\\
E-mail: Georg.Weiglein@durham.ac.uk}

\maketitle

\abstracts{
In the Minimal Supersymmetric Standard Model we present the calculation
of the single charged Higgs-boson production in the $\ga W$- or
$ZW$-fusion and the charged Higgs strahlung channel, \eeenH. The 
set of all \order{\al} corrections arising from loops of Standard
Model fermions and
scalar fermions are taken into account.  Contrary to the case of single
neutral heavy CP-even Higgs-boson production, for the charged Higgs
boson we find for all the parameter space of the typical benchmark
scenarios a cross section smaller than $\sim 0.01 \fb$ for 
$\sqrt{s}/2 \lesssim \MHp$. 
}


\section{Introduction}

Disentangling the mechanism that controls electroweak symmetry
breaking is one 
of the main tasks of the current and next generation of colliders. The
prime candidates are the Higgs mechanism within the Standard Model (SM) or
within the Minimal Supersymmetric Standard Model (MSSM). Contrary to
the SM, two Higgs doublets are required in the MSSM, resulting in five
physical Higgs bosons: the light and heavy CP-even $h$ and $H$, the
CP-odd $A$, and the charged Higgs bosons $H^\pm$.
The Higgs sector of the MSSM can be expressed at lowest
order in terms of $\MZ$, $\MA$, and $\tb = v_2/v_1$, the ratio of the
two vacuum expectation values. 

Pair production of the heavy MSSM Higgs bosons at a Linear Collider
(LC) is limited by kinematics
to $\MH \approx \MA \approx \MHp \lesssim \sqrt{s}/2$.  Thus single
Higgs-boson production (including electroweak loop effects) has recently
drawn considerable interest in the literature
\cite{eennH,other}.
It has been found that the processes \eennH\ could allow for the discovery
of the $H$ significantly beyond the kinematic limit once the dominant loop
corrections are taken into account \cite{eennH}.  On the other hand,
$e^+e^- \to \nu_e \bar \nu_e A$, 
$e^+e^- \to Z^* \to H\{Z,A\}$, 
$e^+e^- \to W^\pm H^\mp$, 
and $e^+e^- \to t\bar b H^-$~\cite{other} 
only possess a small potential to produce the
heavy MSSM Higgs bosons with $\MH \approx \MA \approx \MHp > \sqrt{s}/2$.

Here we present results for the channel \eeenH\ in the MSSM.
Since there is no tree-level~$\{\ga,Z\}W^\pm H^\mp$ coupling, the single
charged-Higgs production starts at the one-loop level.  We take into
account the leading corrections arising from the full set of SM fermion and
sfermion loops. In the case of \eennH\ this type of diagrams
constitutes the leading contribution affecting the decoupling behavior
of the $H$~\cite{eennH}. 


\section{Calculation}

The \onel\ SM fermion and sfermion diagrams for the process \eeenH\ 
are generically depicted in \reffi{fig:FD_eeenH} ($s$-channel diagrams
%
\begin{figure}[ht]
\vspace{-1em}
\unitlength=1bp%
\begin{feynartspicture}(324,100)(3,1)
\FADiagram{}
\FAProp(0.,15.)(9.,16.)(0.,){/Straight}{-1}
\FALabel(4.32883,16.5605)[b]{$e$}
\FAProp(0.,5.)(9.,4.)(0.,){/Straight}{1}
\FALabel(4.32883,3.43948)[t]{$e$}
\FAProp(20.,17.)(9.,16.)(0.,){/Straight}{1}
\FALabel(14.3597,17.5636)[b]{$e$}
\FAProp(20.,10.)(15.,10.)(0.,){/ScalarDash}{-1}
\FALabel(17.5,11.07)[b]{$H$}
\FAProp(20.,3.)(9.,4.)(0.,){/Straight}{-1}
\FALabel(14.3597,2.43637)[t]{$\nu_e$}
\FAProp(9.,16.)(9.,13.5)(0.,){/Sine}{0}
\FALabel(7.93,14)[r]{$\gamma,Z$}
\FAProp(9.,4.)(9.,6.5)(0.,){/Sine}{1}
\FALabel(7.93,5.5)[r]{$W$}
\FAProp(15.,10.)(9.,13.5)(0.,){/Straight}{-1}
\FALabel(12.301,12.6089)[bl]{$f$}
\FAProp(15.,10.)(9.,6.5)(0.,){/Straight}{1}
\FALabel(12.301,7.39114)[tl]{$f'$}
\FAProp(9.,13.5)(9.,6.5)(0.,){/Straight}{-1}
\FALabel(7.93,10.)[r]{$f$}
\FAVert(9.,16.){0}
\FAVert(9.,4.){0}
\FAVert(15.,10.){0}
\FAVert(9.,13.5){0}
\FAVert(9.,6.5){0}

\FADiagram{}
\FAProp(0.,15.)(9.,16.)(0.,){/Straight}{-1}
\FALabel(4.32883,16.5605)[b]{$e$}
\FAProp(0.,5.)(9.,4.)(0.,){/Straight}{1}
\FALabel(4.32883,3.43948)[t]{$e$}
\FAProp(20.,17.)(9.,16.)(0.,){/Straight}{1}
\FALabel(14.3597,17.5636)[b]{$e$}
\FAProp(20.,10.)(15.,10.)(0.,){/ScalarDash}{-1}
\FALabel(17.5,11.07)[b]{$H$}
\FAProp(20.,3.)(9.,4.)(0.,){/Straight}{-1}
\FALabel(14.3597,2.43637)[t]{$\nu_e$}
\FAProp(9.,16.)(9.,13.5)(0.,){/Sine}{0}
\FALabel(7.93,14)[r]{$\gamma,Z$}
\FAProp(9.,4.)(9.,6.5)(0.,){/Sine}{1}
\FALabel(7.93,5.5)[r]{$W$}
\FAProp(15.,10.)(9.,13.5)(0.,){/ScalarDash}{-1}
\FALabel(12.301,12.6089)[bl]{$\tilde f$}
\FAProp(15.,10.)(9.,6.5)(0.,){/ScalarDash}{1}
\FALabel(12.301,7.39114)[tl]{$\tilde f'$}
\FAProp(9.,13.5)(9.,6.5)(0.,){/ScalarDash}{-1}
\FALabel(7.93,10.)[r]{$\tilde f$}
\FAVert(9.,16.){0}
\FAVert(9.,4.){0}
\FAVert(15.,10.){0}
\FAVert(9.,13.5){0}
\FAVert(9.,6.5){0}

\FADiagram{}
\FAProp(0.,15.)(10.,16.)(0.,){/Straight}{-1}
\FALabel(4.84577,16.5623)[b]{$e$}
\FAProp(0.,5.)(10.,4.)(0.,){/Straight}{1}
\FALabel(4.84577,3.43769)[t]{$e$}
\FAProp(20.,17.)(10.,16.)(0.,){/Straight}{1}
\FALabel(14.8458,17.5623)[b]{$e$}
\FAProp(20.,10.)(15.5,10.)(0.,){/ScalarDash}{-1}
\FALabel(17.75,11.07)[b]{$H$}
\FAProp(20.,3.)(10.,4.)(0.,){/Straight}{-1}
\FALabel(14.8458,2.43769)[t]{$\nu_e$}
\FAProp(10.,16.)(10.,10.)(0.,){/Sine}{0}
\FALabel(8.93,13.)[r]{$\gamma,Z$}
\FAProp(10.,4.)(10.,10.)(0.,){/Sine}{1}
\FALabel(8.93,7.)[r]{$W$}
\FAProp(15.5,10.)(10.,10.)(0.8,){/ScalarDash}{1}
\FALabel(12.75,13.27)[b]{$\tilde f'$}
\FAProp(15.5,10.)(10.,10.)(-0.8,){/ScalarDash}{-1}
\FALabel(12.75,7.2)[t]{$\tilde f$}
\FAVert(10.,16.){0}
\FAVert(10.,4.){0}
\FAVert(15.5,10.){0}
\FAVert(10.,10.){0}

\end{feynartspicture}
\vspace{-2.5em}
\caption{Generic $t$-channel diagrams for the process \eeenH.
\label{fig:FD_eeenH}}
\vspace{-1.0em}
\end{figure}
%
and self-energy corrections have been omitted). The contributions
involve all corrections from SM fermion and sfermion loops (which give
contributions only to self-energies and vertices). 
Contributions $\propto m_e/\MW$ were neglected.\\
Furthermore, counterterm contributions are needed for the $W^\pm H^\mp$
self-energy corrections, see \citere{WHren}.  
In order to generate the counterterms, 
it is sufficient to introduce the field renormalization constant for
the $H^\pm - W^\pm$ mixing, $\de Z_{HW}$. This leads to the
Feynman rules for the counterterms, 
\begin{align}
\label{ct-wh}
\Gamma_{\text{CT}}[H^\mp W^\pm(k^\mu)] & = 
        i \frac{k^\mu}{\MW} \MW^2\, \delta Z_{HW}\;,\\
\Gamma_{\text{CT}}[\gamma_\mu W^\pm_\nu H^\mp] & = 
        - i e \MW g_{\mu\nu}\,\delta Z_{HW}\;,\\
\label{ct-zwh}
\Gamma_{\text{CT}}[Z_\mu W^\pm_\nu H^\mp] & =
          i e \MW \frac{s_w}{c_w} g_{\mu\nu}\, \delta Z_{HW}~.
\end{align}
In the on-shell scheme $\de Z_{HW}$ is given by
\begin{align}
\de Z_{HW} & = 1/\MW^2\,\limfunc{Re}\Si_{HW}(\MHp^2)~.
\end{align}
The Feynman diagrams were generated and evaluated with the packages 
{\em FeynArts}, {\em FormCalc}, and {\em LoopTools} 
\cite{feynarts,formcalc,fa-fc-lt}.


\section{Results}

\begin{figure}[ht]
\centerline{\epsfxsize=4.1in\epsfbox{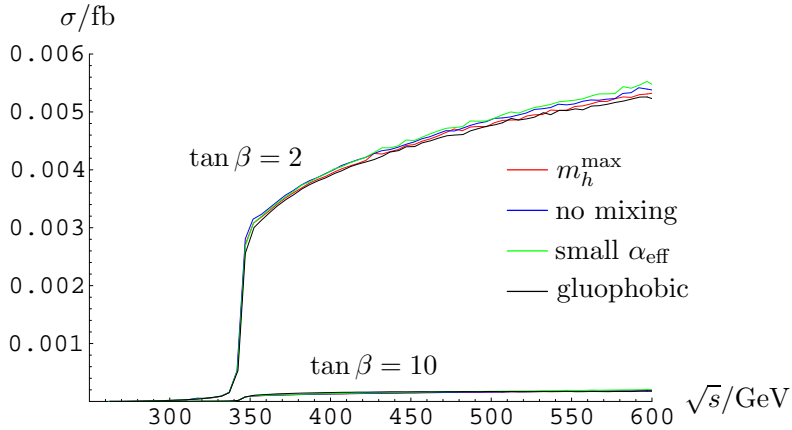}}
\caption{The \eeenHm\ production cross section as a function of $\sqrt{s}$.
\label{fig:results}}
\end{figure}

The results for $H^+$ and $H^-$ production are the same if CP is not
violated (which we assume throughout the paper).  In \reffi{fig:results}
we show the typical size of the production cross section for \eeenHm\
for unpolarized external particles.  The parameters are chosen according
to the four benchmark scenarios described in \citere{LHbenchmark}, with
$\MA = 250 \gev$ and $\tb = 2$ and~$10$ (with $\MHp \approx 262 \gev$). 
Concerning the discovery of the charged Higgs boson at a LC, the
number of expected events is obtained from a twice as large cross
section, due to the production of both, $H^+$ and $H^-$.
%
%
\begin{wrapfigure}[4]{r}{80bp}
\vspace*{-6ex}
\begin{feynartspicture}(80,80)(1,1)   
\FADiagram{}
\FAProp(0.,15.)(5.5,10.)(0.,){/Straight}{-1}
\FALabel(2.18736,11.8331)[tr]{$e$}
\FAProp(0.,5.)(5.5,10.)(0.,){/Straight}{1}
\FALabel(3.31264,6.83309)[tl]{$e$}
\FAProp(20.,17.)(15.5,13.5)(0.,){/Straight}{1}
\FALabel(17.2784,15.9935)[br]{$e$}
\FAProp(20.,10.)(15.5,13.5)(0.,){/Straight}{-1}
\FALabel(18.2216,12.4935)[bl]{$\nu_e$}
\FAProp(20.,3.)(12.,10.)(0.,){/ScalarDash}{-1}
\FALabel(15.4593,5.81351)[tr]{$H$}
\FAProp(5.5,10.)(12.,10.)(0.,){/Sine}{0}
\FALabel(8.25,8.93)[t]{$\gamma,Z$}
\FAProp(15.5,13.5)(12.,10.)(0.,){/Sine}{0}
\FALabel(13.134,12.366)[br]{$W$}
\FAVert(5.5,10.){0}
\FAVert(15.5,13.5){0}
\FAVert(12.,10.){-1}
\vspace*{-5ex}
\end{feynartspicture}
\end{wrapfigure}
In \reffi{fig:results} the cross section for \eeenHm\ is shown as a
function of $\sqrt{s}$.  
The rise of the cross section at $\sqrt{s} \approx \MHp + \MW$ is due
to the $W$ propagator in the type of diagram on the right becoming 
resonant.  The resonance was treated with a fixed width.

%
%
The variation within the four benchmark scenarios is small.  For 
$\tb = 10$ the charged-Higgs production cross section stays at a
negligible level.  Even for $\tb = 2$ it stays below $0.01 \fb$ for 
$2\,\MHp \approx \sqrt{s} \lesssim 500 \gev$.  Using polarized 
$e^+$ and $e^-$ beams, the cross section could be enhanced by about a
factor of~2.   

In summary, the single charged-Higgs production, \eeenH, is most relevant
for small values of $\tb$, which are still marginally allowed from LEP
Higgs searches if the experimental error on the top mass and
theoretical uncertainties are taken into
account~\cite{tbexcl,mhiggsAEC}. 
This process could
possibly increase the potential of a LC for the detection of the heavy
MSSM Higgs-boson spectrum only for parameters beyond the typical
benchmark scenarios.


\end{document}